\documentclass[traditabstract]{aa}  
\usepackage{graphicx}
\usepackage{txfonts}

\def\Msun{\hbox{$\rm\thinspace M_{\odot}$}}
\begin{document}
   \title{PHL\,5038: A spatially resolved white dwarf $+$ brown dwarf binary}


   \author{P.R. Steele\inst{1}, M.R. Burleigh\inst{1}, J. Farihi\inst{1}, B.T. G\"ansicke\inst{2}, R.F. Jameson\inst{1}, P.D. Dobbie\inst{3} \and M.A. Barstow\inst{1}
          }

   \institute{Department of Physics \& Astronomy, University of Leicester,
              Leicester, LE1 7RH, UK\\
              \email{prs15@star.le.ac.uk}
         \and
             Department of Physics, University of Warwick, Coventry,
	     CV4 7AL, UK
	 \and
	     Anglo-Australian Observatory, PO Box 296, Epping, Sydney, NSW 1710, Australia\\
             }

   \date{Submitted January 21, 2009, Accepted March 20, 2009}

 
  \abstract{
    A near-infrared excess is detected at the white dwarf PHL\,5038 in UKIDSS photometry, consistent with the presence of a
   cool, substellar companion. We have obtained $H$- and $K$-grism spectra and images of PHL\,5038 using NIRI on Gemini North. The target is spatially and spectrally resolved into two components; an 8000\,K DA white dwarf, and a likely L8 brown dwarf companion, separated by $0.94\arcsec$. The spectral type of the secondary was determined using standard spectral indices for late L and T dwarfs. The projected orbital separation of the binary is 55\,AU, and so it becomes only the second known wide WD$+$dL binary to be found after GD\,165AB. This object could potentially be used as a benchmark for testing substellar evolutionary models at intermediate to older ages.
  }  

   \authorrunning{Steele et al.}
   \titlerunning{A spatially resolved white dwarf $+$ brown dwarf binary}


   \maketitle


\section{Introduction}

The study of low mass companions to white dwarfs (WDs) probes high mass ratio binary formation and evolution, including the known deficit of brown dwarf (BD) companions to main-sequence stars  (McCarthy \& Zuckerman 2005; Grether \& Lineweaver 2006) . There are currently no empirical constraints on substellar evolutionary models at intermediate to older ages ($>1$Gyr), such as those expected for most WDs. Thus wide, detached substellar companions to WDs can provide benchmarks for testing BD cooling models at these ages (Pinfield et al. 2006). Furthermore, the Earth-sized radii of WDs provide excellent contrast for the infrared detection of these cool, self luminous companions (Probst 1983a, b; Zuckerman \& Becklin 1987; Burleigh, Clarke \& Hodgkin 2002; Farihi, Becklin \& Zuckerman 2008).\\
\indent However,  BD companions to WDs are rare ($<0.5\%$; Farihi, Becklin \& Zuckerman 2005). So far only 3 L-type companions to WDs have been discovered; GD165B (L4, Becklin \& Zuckerman 1988; Kirkpatrick et al. 1999), GD1400\,B (L6-L7, Farihi \& Christopher 2004; Dobbie et al. 2005; Burleigh et al. 2009) and WD\,0137$-$349\,B (L8, Maxted et al. 2006; Burleigh et al. 2006). Of these, GD165 is the only spatially resolved widely separated system.\\
\indent Here we report a near-infrared (NIR) excess at the WD PHL 5038. Follow up observations reveal a spatially resolved substellar companion of late L-type; only the second such pair found (over 20 years later) after GD 165AB. 

\begin{figure}
\centering
\includegraphics[width=8.7cm]{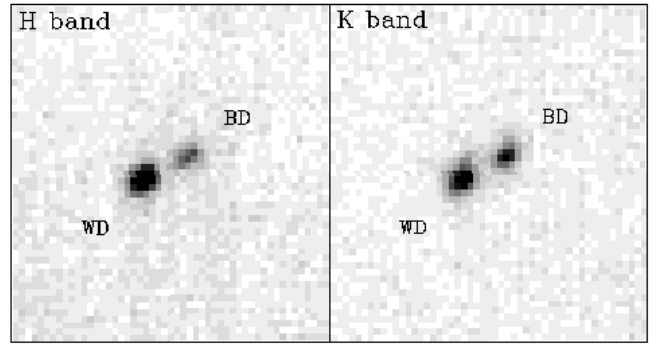} 
\caption{Near-infrared acquisition images of PHL\,5038 taken with Gemini $+$ NIRI immediately prior to $H$- and $K$-grism spectroscopy.  Frames are $6\arcsec$ across with north up and east left.  The brown dwarf companion, PHL\,5038B, is located $0.94\arcsec$ from A at position angle $293.2\arcsec$, as measured in the $K$-band image shown above.}
\label{images}
\end{figure}

\section{Target Identification}

PHL\,5038 was first reported by Haro \& Luyten (1962) in a photographic search for faint blue stars near the South Galactic Pole with $B = 18$~mag, and a poorly constrained position of $22^{\rm h}20.5^{\rm m} -00^{\circ} 43'$ (J2000). This star was recovered in the Sloan Digital Sky Survey as SDSS\,J\,222030.68$-$004107.3 (Eisenstein et al. 2006) , a WD with $T_{\rm eff}\approx8037$\,K and log\,$g=8.28$, yielding an accurate position for the WD.  

\begin{table*}
\caption{Near-Infrared photometry of the PHL\,5038 binary system and nearby calibration star.}
\label{mags}
\begin{center}   
\begin{tabular}{c c c c c c}        
\hline\hline  
Bandpass &      $\lambda$     &  PHL\,5038 &       PHL\,5038A  &    PHL\,5038B    &          2MASS J22203205-0040210\\
         &      ($\mu$m)      &  (mag) &      (mag)      &    (mag)        &          (mag)\\
\hline
$Y^{1}$      &      1.03        & $16.89\pm0.01$ &   -   &   -     & - \\
$J^{1}$      &      1.25        & $16.73\pm0.02$ &   -   &   -     & - \\
$H^{1}$      &      1.63        & $16.44\pm0.03$ &   -   &   -     & - \\
$K^{1}$      &      2.20        & $16.31\pm0.04$ &   -   &   -     & - \\
$H^{2}$      &      1.65        &   -   &     $16.82\pm0.07$ & $17.84\pm0.08$    &      $14.447\pm0.053$\\
$K^{2}$      &      2.19        &   -   &     $16.72\pm0.07$ & $17.18\pm0.08$    &      $14.161\pm0.051$\\
\hline
\end{tabular}
\end{center}
$^{1}$UKIDSS passbands as described in Hewett et al. (2006).
$^{2}$$H$- and $K$-band photometry were obtained on a single night.  The photometric
errors were calculated using a 5\% error from the nearby 2MASS star used
for flux calibration, a $2-3$\% photometric measurement error, and an
assumed 5\% internal error typical of near-infrared arrays.  The final
column displays the nearby 2MASS calibration star with its 2MASS catalogue
photometry and errors. Colour transformations between the filters are of the order of 1\% (Carpenter 2001) and so are neglected.
\end{table*}

The WD was detected in the UKIDSS Large Area Survey (LAS), with DR4 photometry for  PHL\,5038 listed in Table~1. These values give an excess of $7-10$\,$\sigma$  when compared to the predicted values for the WD of $(J,H,K)=(16.86,16.72,16.72)$\,mag, calculated by folding our WD model (Section~4.3) through the appropriate UKIDSS filter response curves.

The UKIDSS data flag (mergedClass) indicated the possibility that PHL\,5038 was an extended source at one or more bandpasses. Follow-up observations were warranted to investigate the nature of the excess flux and its possible spatially extended nature . Thus, this object was included in a Gemini programme to clarify the nature of NIR excesses to WDs in the UKIDSS LAS.

\section{Observations}

Near-infrared grism spectroscopy of PHL\,5038 was obtained at Gemini North using NIRI 
(\cite{hod03}).  During the first acquisition images, the science target was spatially 
resolved into two components separated by  $0.94\arcsec$, and the slit was placed
over both point sources for all subsequent programme observations.  On 24 Aug 2008, a 
single 30 s $H$-band acquisition image was obtained prior to placement of the science 
targets on the slit.  Spectroscopy was performed with the $H$ grism and the 6 pixel 
slit (covering $1.43-1.96$ $\mu$m at $R\approx500$) with eight, single coadd frames 
of 225 s each at one of two offset positions along the slit, for a total exposure 
time of 30 min.  The A0V telluric standard HIP\,102631 was observed with 
eight frames of 1s and three coadds.  Internal flat field and arc lamp calibration 
frames were taken immediately following the science target spectroscopy.

Three attempts were made to obtain $K$-band spectroscopy of PHL 5038.  On 12 Aug 
2008, the science target was imaged for 30 s at two offset positions, then placed on 
the slit, but the spectroscopic sequence was abandoned before usable data were taken. However, the acquisition image from this date could still be used for photometry. 
On 21 Aug 2008, the target was also imaged prior to being placed on the slit.  
Spectroscopy was performed with the $K$ grism and the 6 pixel slit (covering $1.90-
2.49$ $\mu$m at $R\approx500$), with six, single coadd frames of 225 s each at one 
of two offset positions along the slit, for a total integration time of 22.5 min. 
The A0V telluric standard HIP\,102631 was observed with eight frames of 1 
s and three coadds.  Internal flat field and arc lamp calibration frames were taken 
immediately following the science target spectroscopy.  Unfortunately, the fainter of 
the two point sources was poorly aligned with the slit, resulting in a significant 
loss of spectroscopic signal for that component (see below).  On 6 Nov 2008, an 
identical $K$-band imaging sequence preceded a successful placement of both point 
sources on the slit. However, due to poor seeing the system was not spatially well resolved, and the S/N was too low to merit an extraction for PHL\,5038B.

\section{Data and Analysis}

All NIRI frames were processed using Gemini IRAF {\sf niri} and {\sf gnirs} 
packages, including the creation of appropriate flat-field images, bad pixel masks, 
and wavelength calibrated spectral data.  

\subsection{Photometry}

Since only a single $H$-band image was
obtained, this frame was cleaned of bad pixels and flat fielded.  For the $K$-band
image pairs, the frames were cleaned of bad pixels, flat fielded, pair subtracted, 
shifted and recombined.  Fully reduced $H$- and $K$-band images are shown in Figure 
1; the WD is designated PHL\,5038A and the cool companion is PHL\,5038B.

Aperture photometry and astrometry on these images was performed with the IRAF task 
{\sf apphot} using aperture radii of 2, 3, and 4 pixels ($0.117\arcsec$ pixel$^{-1}$) and 
a sky annulus of $16-20$ pixels.  For most frames, photometry obtained with the 4-pixel
aperture radius was discarded owing to probable contamination from the other binary
component.  A single $H$- and $K$-band measurement was made for both PHL\,5038 A and B.  The science target photometry
was calibrated using a nearby ($52\arcsec$ distant) field star, 2MASS J22203205-0040210, 
which fell onto the NIRI array in all imaging observations.  Photometric data for
PHL\,5038 A and B, as well as the calibrator are given in Table 1. 

\subsection{Spectroscopy}

\begin{figure*}
\centering
\label{spectrum1}
\includegraphics[width=4.9cm,angle=270]{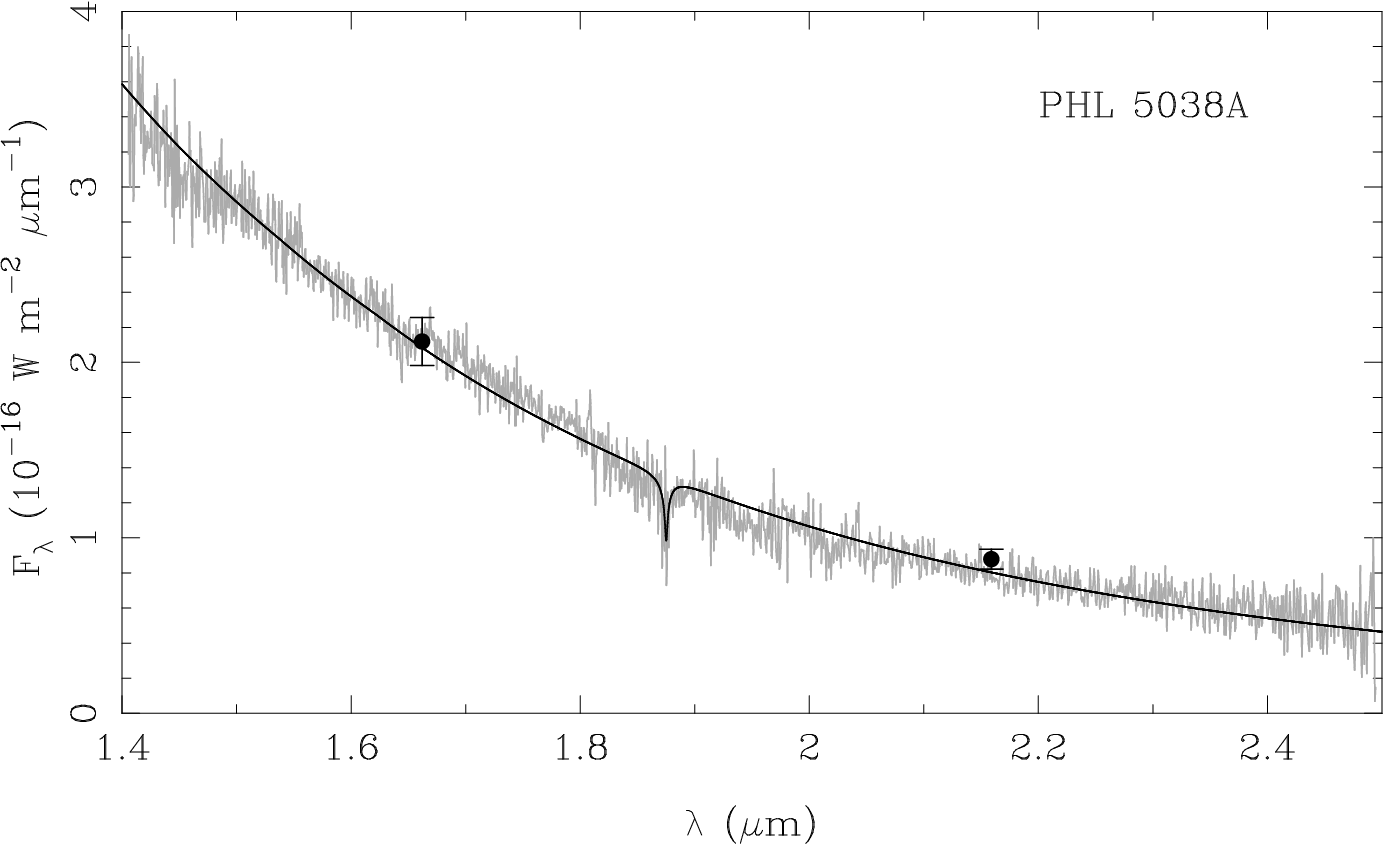}
\caption{Modelled (black) and observed (grey) $HK$ spectra of PHL\,5038\,A plotted with photometry as measured in the acquisition images. Note the detection of Pa$_{\alpha}$ absorption.}

\centering
\label{spectrum2}
\includegraphics[width=4.9cm,angle=270]{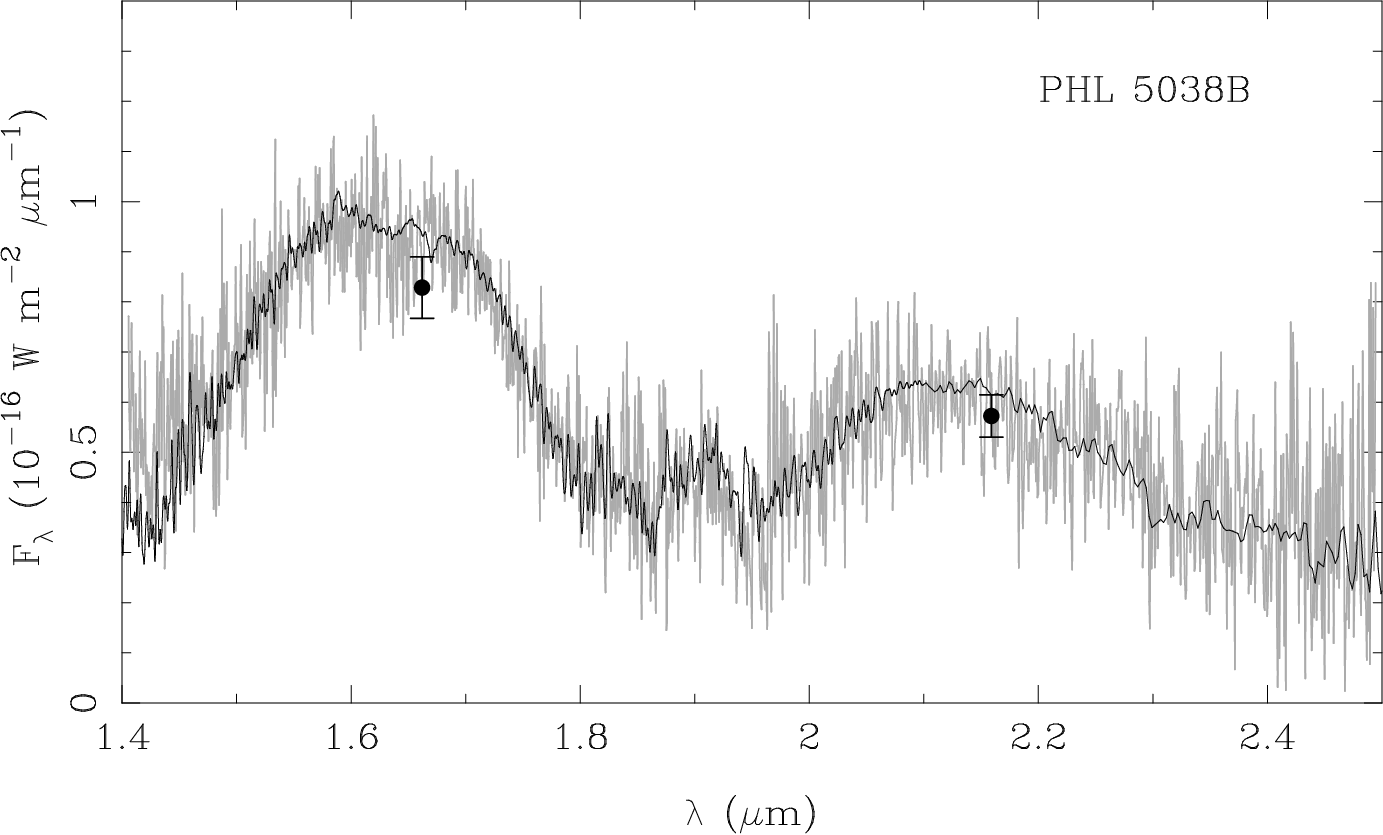}
\caption{Observed $HK$ spectrum (grey) of PHL\,5038\,B plotted with photometry as measured in the acquisition images. An L8 spectrum (2MASS\,1632$+$1904) has been scaled to match the flux of the secondary and over-plotted (black) for comparison.}
\end{figure*}

The spectroscopic frames were cleaned of bad pixels, flat fielded, pairwise subtracted,
shifted, combined, and wavelength calibrated.  A slightly more aggressive (relative to 
the default {\sf nscombine} settings) outlier rejection algorithm was applied in the 
next to last step, which produced visibly superior spectral images for extraction.  In 
the $H$-grism observations, the spectra of PHL\,5038 A and B were well resolved spatially, 
and each was extracted using a 2-pixel spectral aperture. For the $K$-grism spectral extraction, we only used the data from 21 August 2008. In these observations the two science target spectra were spatially resolved, but the 
signal of component B was significantly fainter than expected due to poor slit alignment.
The data are still usable and a spectrum of each component was extracted using a 2-pixel 
spectral aperture, but care had to be taken with PHL\,5038B, and the default aperture 
function had to be carefully corrected to avoid automatically picking up the signal of 
the WD. The telluric standard data was reduced and extracted identically.

Flux calibration and telluric feature correction were performed for each order with 
Spextool (version 3.3; Cushing, Vacca \& Rayner 2004; Vacca, Cushing \& Rayner 2003), using the general {\sf xtellcor} package.
Although in principle this should produce a well calibrated spectrum, there are a couple of
reasons why this proved difficult or impossible in this case: 1) the telluric standard and science observations differed by 0.2 airmasses, and 2) slit losses; both ordinary and those caused by clear star-slit misalignment.  Therefore 
the spectral data were calibrated from the Table 1 photometry; some inter-order adjustment 
was necessary for both the WD and the companion.

The spectral data were further processed to interpolate over data points where residual 
bad pixels or cosmic rays persisted (around several to one dozen), to crop the noise
dominated edges, and to merge the data into a single, contiguous spectrum for each binary
component.  The fully processed spectra for PHL\,5038A and B are shown in Figures 2 and 3 respectively.

\subsection{Spectral Analysis}

\begin{table*}
\caption{Spectral indices as measured for PHL\,5038\,B, L8, L9 and T0 fields dwarfs (Burgasser et al. 2006).} 
\label{specindex} 
\centering   
\begin{tabular}{c c c c c}    
\hline\hline    
Index & PHL\,5038\,B & 2MASS\,1632$+$1904 (L8) & DENIS\,0255$-$4700 (L9) & SDSS\,0423$-$0414 (T0) \\   
\hline                    
H$_{2}$0-$H$ & $0.74\pm0.08$ & 0.71 & 0.69 & 0.64 \\
CH$_{4}$-$H$ & $0.98\pm0.08$ & 1.08 & 1.03 & 0.99 \\
H$_{2}$0-$K$ & $0.76\pm0.14$ & 0.70 & 0.69 & 0.68 \\
CH$_{4}$-$K$ & $0.89\pm0.14$ & 0.88 & 0.84 & 0.82 \\
\hline                             
\end{tabular}
\end{table*}
An atmospheric fit of the SDSS spectrum of PHL\,5038A using the models of Koester et al. (2005) yielded values of $T_{\rm eff}=8000\pm100$\,K and log\,$g=8.2\pm0.2$, implying a mass of $M_{\rm WD}=0.72\pm0.15$\Msun \ and a distance of $64\pm10$\,pc. The model spectrum is plotted over the observed spectrum in Figure~2 and is consistent with the detection of Pa$_{\alpha}$ absorption. No contamination from the secondary can be seen indicating we have successfully separated the components in our spectral analysis.\\
\indent To determine the spectral type of PHL 5038B, we measured the standard spectral indices for late L dwarfs and T dwarfs (Burgasser et al. 2006), and listed them in Table~2 with established L8, L9 and T0 indices for comparison.  The errors were determined by taking the standard deviation over the spectral index denominators for the $H$- and $K$-band regions. These values indicate the spectral type is most likely L8-L9, with the H$_{2}$O$-H$ index suggesting a spectral type of T0 is less likely .\\ 
\indent The observed $HK$ magnitudes (Table~1) are also consistent with a late L-type BD at a distance of 64\,pc.  Although the S/N of the spectrum is insufficient for the detection of any distinct absorption lines, the overall shape of the spectrum is also in best agreement with a spectral type of L8. The spectrum of an L8 field BD (2MASS$1632+1904$, McLean et al. 2003) has been scaled to 64\,pc and plotted over the observed spectrum of PHL\,5038B in Figure~3 for direct comparison. 

\section{Discussion}
Caballero, Burgasser \& Klement (2008) list the spatial density of L8-L9 field dwarfs as $0.64\times10^{-3}\rm\,pc^{-3}$. It follows that the number of expected L8-L9 field dwarfs within a cylinder of angular radius $=1\arcsec$ and length $=20$\,pc centered on PHL\,5038A is $\approx3\times10^{-9}$. Therefore, PHL\,5038A and B are almost certainly a physical pair.\\
\indent The $0.94\arcsec$ angular separation of the binary, at  the estimated distance of 64\,pc equates to a projected orbital separation of $55$\,AU. When the main-sequence progenitor of PHL\,5038A evolved into a WD the orbital separation of PHL\,5038B expanded by a maximum factor of $ \rm M_{MS}/M_{WD}$ (Jeans 1924; Zuckerman \& Becklin 1987; Burleigh, Clarke \& Hodgkin 2002). For $\rm M_{WD}=$0.72\,\Msun\ and $\rm M_{MS}=$3.2\,\Msun, this factor would have been  $\approx4.4$. Thus, the original projected semi-major axis was $>13$\,AU.\\
\begin{table}
\caption{MS progenitor masses (M$_{\rm MS}$) and ages (t$_{\rm MS}$), WD cooling ages (t$_{\rm C}$), total age (t$_{total}$), and MS age as a fraction of WD cooling age for PHL\,5038A. } 
\label{param} 
\centering   
\begin{tabular}{c c c c c}
\hline    
  & & M$_{\rm WD}$ (\Msun) &  \\                  
 & 0.57 & 0.72 & 0.87 \\     
\hline
M$_{\rm MS}$ (\Msun) & 2.1 & 3.2 & 4.4 \\ 
t$_{\rm MS}$ (Gyr) & 1.3 & 0.4 & 0.2 \\
t$_{\rm C}$ (Gyr) & 1.0 & 1.5 & 2.5 \\
t$_{total}$ (Gyr) & 2.3 & 1.9 & 2.7  \\
t$_{\rm MS}/t_{\rm C}$ & 1.3 & 0.3 & 0.1 \\
\hline                             
\end{tabular}
\end{table}
\indent The only other resolved L-type companion to a WD currently known is GD\,165B, which has a projected orbital separation of 120\,AU. Whether or not the secondary in this system is a true BD is still a matter of some debate (Zuckerman \& Becklin 1992; Kirkpatrick, Henry \& Liebert 1993; Kirkpatrick et al. 1999; Zuckerman et al. 2003). We estimate the temperature of PHL\,5038B to be $1400-1500$\,K by comparison to L-type field dwarfs with measured effective temperatures (Vrba et al. 2004). At an age of 1.9-2.7\,Gyr (Table~3) the Lyon group models predict a mass of around 60\,M$_{\rm J}$ (Chabrier et al. 2000; Baraffe et al. 2001). Therefore, PHL\,5038B is almost certainly a bona fide substellar object.\\
\indent The mass luminosity relation for BDs is age dependent. Benchmark BDs can be used to test evolutionary models for substellar objects if their ages can be satisfactorily constrained (e.g., Dupuy et al. 2009). One of the candidates for these studies is widely orbiting WD$+$BD binaries (Farihi, Becklin \& Zuckerman 2005; Pinfield et al. 2006) such as PHL\,5038. The limiting factor in this system is its age calibration. Pinfield et al. (2006) require that a good age calibrator's progenitor lifetime be no more than 10\% of the WD cooling age (t$_{\rm MS}/t_{\rm C}$), due to intrinsic uncertainty in calculating main sequence (MS) lifetimes.\\
\indent With $T_{\rm eff}=8000\pm100$\,K and log\,$g=8.2\pm0.1$ the cooling age of PHL\,5038A is $1.0-2.5$\,Gyr (Fontaine, Brassard \& Bergeron 2001). The mass of the progenitor to PHL\,5038A is estimated to be 3.2$\pm1.1$\,\Msun\ using the initial to final mass relationship of Dobbie et al. (2006), which is valid for initial masses of $\ge1.6$\Msun\ (Kalirai et al. 2007). This indicates the WD descended from an A or late B-type star. An approximate main sequence lifetime can be calculated using the models of Girardi et al. (2000), giving $0.2-1.3$\,Gyr. Therefore, the total age of the system is $1.9-2.7$\,Gyr. These ages were calculated for the lower and upper mass limits of PHL\,5038A and are listed in Table~3 along with the ratio t$_{\rm MS}/t_{\rm C}$. This ratio ranges from $0.1-1.3$ for a $M_{\rm WD}=0.57-0.87$~\Msun\ and so only fulfills Pinfield's requirement for a benchmark BD if the WD mass is as high as 0.87\,\Msun. At this point, we note that the masses for DA white dwarfs cooler than 12,000\,K derived from spectroscopic fitting techniques may be over-estimated by up to $0.2$\,\Msun, for reasons that are not fully understood (Kepler et al. 2007; Koester et al. 2008).


\section{Conclusions}
We have confirmed the presence of and spectroscopically typed a BD companion to the WD PHL\,5038A. PHL\,5038B was assigned a spectral type of L8-L9 based on the measurement of standard spectral indices in the NIR, the first time this has been accomplished for a bona fide WD$+$BD binary. This spectral type is corroborated by the photometry of PHL\,5038B and by comparison with the spectrum of an L8 field BD. The system is situated $64$\,pc from Earth, and the BD has a projected orbital separation from its primary of $55$\,AU. PHL\,5038B has the potential of being used as a benchmark BD if its ages can be determined more accurately. A measurement of a trigonometric parallax for this system would enable us to more accurately constrain the radius and, via models, the mass of the WD. This would then allow a better estimate for the total age of the system.

\begin{acknowledgements}
PRS is sponsored by STFC in the form of a studentship. MRB acknowledges the support of STFC in the form of an Advanced Fellowship. We thank Detlev Koester for the use of his model atmosphere calculations.
\end{acknowledgements}


\end{document}